\documentclass[onecolumn, doublespace, 12pt fonts]{aastex6}
\usepackage{graphicx,latexsym,amssymb, epsfig}
\usepackage{multirow,amsmath,array,booktabs}
\usepackage{subfigure}
\usepackage{tabularx}
\usepackage{color}
\usepackage{bm}
\usepackage[figuresright]{rotating}
\usepackage[section]{placeins}
\usepackage{slashed}
\usepackage{graphicx}
\usepackage{tikz}

\newcommand{\nc}{\newcommand}       
\nc{\vc}[1] {\mbox{\boldmath $#1$}} 
\nc{\del}       {\partial}              
\nc{\bra}       {\langle}               
\nc{\ket}       {\rangle}               
\nc{\bras}[1]   {\langle #1|}           
\nc{\kets}[1]   {|#1\rangle}            
\nc{\mapleft}[1]{           
	\smash{\mathop{\,          %
			\hbox to 1.5cm{\rightarrowfill}\, }\limits_{#1}}}
\nc{\beq}     {\begin{eqnarray}} \nc{\eeq}    {\end{eqnarray}}
\nc{\nn}      {\\\nonumber} \nc{\vs}      {\vspace{-0.275cm}}
\nc{\fra}    {\frac{1}{2}}
\nc{\mb}        {\mathbf}

\begin{document}
	\title{Properties of neutron star described by a relativistic {\it{ab initio}} model}
	\author{Chencan Wang\altaffilmark{1}, Jinniu Hu,\altaffilmark{1,2}, Ying Zhang\altaffilmark{3,2}, Hong Shen\altaffilmark{1}}
	\altaffiltext{1}{School of Physics, Nankai University, Tianjin 300071,  China}
	\altaffiltext{2}{Strangeness Nuclear Physics Laboratory, RIKEN Nishina Center, Wako, 351-0198, Japan}
	\altaffiltext{3}{Department of Physics, Faculty of Science, Tianjin University, Tianjin 300072, China}

	\email{hujinniu@nankai.edu.cn}

	\begin{abstract}
		Properties of neutron star are investigated by an available relativistic {\it{ab initio}} method, i.e., the relativistic
		Brueckner-Hartree-Fock (RBHF) model, with the latest high-precision relativistic charge-dependent potentials, pvCD-Bonn A,
		B, C. The neutron star matter is solved within the beta equilibrium and charge neutrality conditions in the
		framework of RBHF model. Comparing to the conventional treatment, where the chemical potential of lepton was approximately represented by the symmetry energy of nuclear matter, the equation of state (EOS) of neutron star matter in the present self-consistent calculation with pvCD-Bonn B has striking difference above the baryon number density $n_b=0.55$ fm$^{-3}$. However, these differences influence the	global properties of neutron star only about $1\%\sim2\%$. 	Then, three two-body potentials pvCD-Bonn A, B, C, with different tensor components, are systematically applied in RBHF model to calculate the properties of neutron star. It is found that the maximum masses of neutron star are around $2.21\sim2.30M_\odot$ and the corresponding radii are $R =11.18\sim11.72$ km. The radii of $1.4M_\odot$ neutron star are predicated as $R_{1.4} = 12.34\sim12.91$ km and their dimensionless tidal deformabilities are $\Lambda_{1.4} = 485\sim 626$. Furthermore, the direct URCA process in neutron star cooling will happen from $n_b=0.414\sim0.530$ fm$^{-3}$ with the proton fractions, $Y_p=0.136\sim0.138$. All of the results obtained from RBHF model only with two-body pvCD-Bonn potentials completely satisfy various constraints from recent astronomical observations of massive neutron stars, gravitational wave detection (GW 170817), and mass-radius simultaneous measurement (NICER).
	\end{abstract}
	
	\section{Introduction}
		The pioneering investigator of neutron star can be traced back to Landau, who used to indicate the existence 
		of dense star in universe like gigantic nucleus in 1932~\citep{landau1932}. At that time, Landau was even unaware
		of the discovery of neutron~\citep{chadwick1932}.  The precise concept of neutron star, being the collapsed core during supernova explosion, was proposed by Baade and Zwicky~\citep{baade1934}. Subsequently, the static equilibrium equation to describe the neutron star from the general relativity was derived by Tolman, Oppenheimer and Volkov, i.e.,  Tolman-Oppenheimer-Volkov (TOV) equation~\citep{tolman1939,oppenheimer1939}. Furthermore, the interior compositions of neutron star were also discussed by Hund and Gamow in terms of the beta equilibrium condition in nuclear matter~\citep{hund1936,gamow1937}.  	
		With the developments of observation technology for the X-ray pulse in 1960s~\citep{ciacconi1962}, the radio pulse source, PSR B1919+21, discovered by Bell and Hewish, was firstly confirmed as a neutron star~\citep{hewish1968}. In 1974, the first binary pulsar, PSR B1913+16, was found by Taylor and Hulse, which is composed by two neutron stars. Its orbit is gradually decreasing and perfectly predicted by the general theory of relativity. It is the first circumstantial evidence of gravitational waves existing~\citep{taylor1982}.
		In 2017, the gravitational wave from the binary neutron star merger was firstly detected by advanced LIGO and Virgo collaborations as GW 170817 event~\citep{abbott2017a}, which was jointly confirmed by other astronomical observations later on such as gamma-ray burst and optical transient~\citep{abbott2017b,goldsein2017}. This opens the multi-messenger astronomy era and has a far-reaching effect on astrophysics, nuclear physics, and other subjects.
		
		There have been thousands of neutron star observed since 1967, whose masses are mainly in $1.0\sim2.3M_\odot$ and
		radii are around 10 km~\citep{lattimer2005,lattimer2012,martinez2015}. 
		According to a lot of theoretical investigations, a commonly accepted internal structure of neutron star contains following regions~\citep{lattimer2004}, from the exterior to interior: 
		the atmosphere in the star's surface formed by light elements,
		the outer crust which consists of free electrons and nuclei, 
		the inner crust where the neutron in neutron-rich nuclei beginning to drip out,
		the outer core formed by homogeneous nuclear matter with neutrons, protons, and leptons, and 
		the inner core where the exotic states of baryons and quarks may exist.
		Therefore, the comprehensive understanding of neutron star physics requires the close collaborations between astrophysics and nuclear physics. On the other hand, the observation data of neutron star provides strong constraints on the equation of state (EOS) of nuclear matter at high densities and poses a great challenge to present nuclear structure theories, especially with the precision measurements for the massive neutron stars recently:   
  		PSR J1614-2230 ($1.928\pm0.017M_\odot$)~\citep{demorest2010,fonseca2016}, 
  		PSR J0348+0432 ($2.01\pm0.04M_\odot$)~\citep{antoniadis2013}, and
  		PSR J0740+6620 ($2.14_{-0.09}^{+0.10}M_\odot$)~\citep{cromartie2020}. 
 		In addition, the tidal deformabilities of neutron star, which denotes the deformation of a massive object influenced by an external gravitational field from another massive body, can be extracted from the observables of gravitational wave generated by the binary neutron star merger~\citep{abbott2018, most2018,radice2018,finstad2018}. It provides a new constraint on  the behaviors of nuclear matter at high density region. Furthermore, in 2019, the Neutron star Interior Composition Explorer (NICER) collaboration obtained the first-ever map of neutron star surface about PSR J0030+0451 and measured its mass and radii simultaneously~\citep{raaijimakers2019}. Two independent analysis groups with these data reported a mass of $1.34_{-0.16}^{+0.15}M_\odot$ with a radius of $12.71_{-1.19}^{+1.14}$ km~\citep{riley2019} and a mass of $1.44_{-0.14}^{+0.15}M_\odot$ with a radius of $13.02_{-1.06}^{+1.24}$ km~\citep{miller2019}, respectively.  
  		
  		In TOV equation, the energy density--pressure ($\varepsilon$--$P$) of nuclear matter must be provided by the nuclear many-body theory now, since the available experiments can only constrain the compact matter around $2\sim 3 n_0$, where $n_0$ is the nuclear saturation density, while the central region of neutron star usually approaches $5\sim 8n_0$. The properties of such ultra-dense object can only be obtained by the predictions via the existing nuclear models.  Generally, there are two types of theoretical models, which can derive the reasonable EOSs to describe properties of neutron star properly. The first type is based on nuclear density functional theories with effective nucleon-nucleon ($NN$) interactions,
  		such as Skyrme Hartree-Fock (SHF) model~\citep{vautherin1972, douchin2001, dutra2012}, Gogny Hartree-Fock (GHF) model~\citep{boquera2018}, 
		relativistic mean-field (RMF) model~\citep{shen1998, shen2002, bao2014a,bao2014b}, 
  		relativistic Hartree-Fock (RHF) model~\citep{long2006, long2007, sun2008}, and so on.
  		The effective $NN$ potentials in these models are determined by reproducing the nuclear bulk properties around nuclear saturation density, such as ground-state binding energies and charge radii of finite nuclei, together with the empirical saturation properties of symmetric nuclear matter. Therefore, the density functional theories have large uncertainties when their EOSs are extrapolated to high-density region. 
  		
  		The other type is based on the {\it ab~initio} many-body models with the realistic $NN$ interaction which is obtained by fitting the $NN$ scattering data. Most of them were achieved in the non-relativistic framework, such as, Brueckner-Hartree-Fock method~\citep{li06,baldo07,baldo16}, quantum Monte Carlo methods~\citep{akmal98,carlson15}, self-consistent Green's function method~\citep{dickhoff04}, coupled-cluster method~\citep{hagen14a,hagen14}, many-body perturbation theory~\citep{carbone13,carbone14,drischler14}, functional renormalization group (FRG) method~\citep{drews15,drews16}, lowest order constrained variational method~\citep{modarres93}, and so on. These non-relativistic {\it ab~initio} methods can simulate the saturation behavior of symmetric nuclear matter more or less with present high precision realistic $NN$ potentials~\citep{stoks94,wiringa95,machleidt01, entem03,epelbaum05,epelbaum15a,epelbaum15b,entem15,entem17,reinert18}. However, the three-body nucleon force must be included in these models to reproduce the empirical saturation properties of symmetric nuclear matter~\citep{li06,hu17,sammarruca18,logoteta19}.

  		There are also few {\it ab~initio} approaches, which were constructed in relativistic framework, for example, the relativistic Brueckner-Hartree-Fock (RBHF) model. The RBHF model can reasonably describe the saturation properties of symmetric nuclear matter~\citep{brockmann1990} only with two-body nuclear potentials due to the additional repulsive contributions generated from the nucleon-antinucleon excitation ($Z$-diagram). After that, the RBHF model was firstly applied to simulate properties of neutron star by Engvik et al.~\citep{engvik1994,bao1994}, where the decay of neutron in the star was neglected.  
  		Then, the crust structure of neutron star was investigated by Sumiyoshi et al.~\citep{sumiyoshi1995} with the RBHF
  		results in the framework of Thomas-Fermi approximation~\citep{ogasawara1982}, where the neutron star matter with beta equilibrium and charge neutrality conditions was discussed in high density.
  		Further developments have been done by Krastev and Sammarruca, where the integrals about Pauli operator in Bethe-Goldstone equation were exactly treated in asymmetric nuclear matter~\citep{krastev2006,sammarruca2010}.  \added{Later, Katayama and Saito exactly solved the neutron star matter in RBHF model with Bonn potentials by considering beta equilibrium and charge neutrality conditions self-consistently for the first time and investigated the validness of angle-average approximation~\citep{katayama2013}.}
  		
  		In these investigations, there are two essential issues must be improved for the studies of neutron star with RBHF model. First, the Bonn potentials used in the previous calculations cannot describe the charge dependence of $NN$ potential and the latest $NN$ scattering data with high precision. Second, the EOS of neutron star matter in the past was \added{mainly} achieved by the symmetry energy approximation because of the complicated treatments for the asymmetry nuclear matter in RBHF model. \added{Therefore, its validity for properties of neutron star should be discussed in detailed comparing to calculations from the exact method.} Recently, we developed the charge-dependent Bonn potentials with the pseudovector coupling between pion and nucleon (pvCD-Bonn)~\citep{wang19}, based on the original CD-Bonn potential~\citep{machleidt01}, which can describe the $NN$ scattering data with high precision. 
 		Therefore, in this work, the pvCD-Bonn potentials will be applied to study the properties of neutron star in RBHF model, where the neutron star matter with the beta equilibrium and charge neutrality conditions will be solved exactly. The results will be compared to those from the symmetry energy approximation. \added{The latest high-precision relativistic charge-dependent $NN$ potentials will be used and the difference between the symmetry energy approximation and the exact calculation for neutron star matter on properties of neutron star will be clearly obviously exhibited in present work, in comparison with the work by Katayama and Saito in \citet{katayama2013}. }
  		 
  		The contents are arranged as follows: the theoretical frameworks of RBHF model and neutron star matter are reviewed in Sec.~\ref{theorfram}; in Sec.~\ref{results}, EOSs of neutron star matter will be shown. The properties of neutron star with pvCD-Bonn potentials will be discussed. Sec.~\ref{summary} will finally give the summaries and conclusions.
 
	\section{Theoretical framework}\label{theorfram}
		\subsection{Relativistic Brueckner-Hartree-Fock (RBHF) model for nuclear matter}
			In the nuclear medium, the motion of single-nucleon satisfies the Dirac equation with species of isospin $\tau$ ($\tau=p,~n$), 
			\begin{equation}\label{diraceq}
			(\bm{\alpha}\cdot\mb{p}+\beta M_\tau + \beta U_\tau) u_\tau(\mb{p},s)
			=E_\tau u_\tau(\mb{p},s),
			\end{equation}
			where $\mb{p},~s$ denote the momentum and spin of nucleon, respectively. Its solution, $u_\tau(\mb{p},s)$, is 
			a Dirac spinor normalized by $\bar{u}_\tau(\mb{p},s)u_\tau (\mb{p},s)=1$.  
			Due to the transnational and rotational symmetries of infinite nuclear matter, the single-nucleon potential can be approximately decomposed into \citep{brockmann1990},
			\begin{equation}\label{singpot}
			U_\tau\approx U_{\tau,\mathrm{s}}+\beta U_{\tau,\mathrm{v}},
			\end{equation}
			where scalar potential $U_{\tau, \textrm{s}}$ and time component of vector potential $U_{\tau, \textrm{v}}$ are weakly momentum-dependent and are regarded as constants at a fixed baryon density. Furthermore, the spatial components of vector potential are also neglected in this work. \added{Here, it should be emphasized that there are two schemes to treat the self-energy components in RBHF model. The first one is neglecting the momentum dependence of scalar potential and the time component in vector potential, and eliminating the spatial components of the vector potential as we done in this work, which can easily extract the scalar potential and vector potential from the single-nucleon potential. The second way is projecting the effective interaction to five Lorentz covariants and adopting them to calculate the self-energy in relativistic Hartree-Fock framework~\citep{katayama2013}. There, the vector component of vector potential is kept, which provides much fewer contributions to the self-energy comparing to the scalar potential and time component of vector potential as shown in the Fig. 2 of  \citet{katayama2013}. In future, we will clearly discuss its role in neutron star matter.}

			As a result, the potentials $U_{\tau, \textrm{s}}$ and $U_{\tau, \textrm{v}}$  can be absorbed by the effective mass $M_\tau^*$ and effective energy $E_\tau^* $, respectively,
			\begin{equation}\label{effem}
			M_\tau^* = M_\tau + U_{\tau,\mathrm{s}},  \quad 
			E_\tau^* = E_\tau-U_{\tau,\mathrm{v}}. 
			\end{equation}
			Therefore, the Dirac equation in the nuclear medium, i.e. Eq~\eqref{diraceq} can be rewritten as
			\begin{equation}\label{ImDiracEq}
			(\bm{\alpha}\cdot\mb{p}+\beta M_\tau^*) u_\tau(\mb{p},s)=E_\tau^* u(\mb{p},s)
			\end{equation}
			with a plane wave solution $u_\tau(\mb{p},s)$,
			\begin{equation}\label{ImDiracEq-Solut}
			E_\tau^*=\sqrt{p^2+M_\tau^{*2}}, \qquad u_\tau(\mb{p},s) =
			\sqrt{\frac{E_\tau^*+M_\tau^*}{2M_\tau^*}} \left(\begin{array}{c}
			1 \\
			\frac{\bm{\sigma}\cdot \mb{p}}{E_\tau^*+M_\tau^*}
			\end{array}\right).
			\end{equation}
			
			In the mean-field approximation,  the single-nucleon potential  $U_\tau$ represents the average interaction of one nucleon generated by other nucleons . Because of the medium effect, the realistic $NN$ potential
			$V_{\tau_1\tau_2}(\mathbf{q}',\mathbf{q})$ should be replaced by effective 
			$G$ matrices in nuclear many-body system in the RBHF model, which can be obtained by solving the in-medium Blankenbecler-Sugar (BbS) equation. It was reduced from the Bethe-Salpter equation,
			\begin{equation}\label{BbS-Gmatrix}
			G_{\tau_1\tau_2}(\mathbf{q}',\mathbf{q}|\mathbf{P}) = V_{\tau_1\tau_2}
			(\mathbf{q}',\mathbf{q}) + \int \frac{d^3k}{(2\pi)^3}V_{\tau_1\tau_2}
			(\mathbf{q}',\mathbf{k})\frac{2W_k}{W_0+W_k}
			\frac{Q_{\tau_1\tau_2}(\mathbf{k},\mathbf{P})}{W_0 - W_k}
			G_{\tau_1\tau_2}(\mathbf{k},\mathbf{q}|\mathbf{P}),
			\end{equation}
			where $\mathbf{q}'$, $\mathbf{k}$, $\mathbf{q}$ denote the initial, intermediate,
			and final relative momenta, respectively. $W_{q'}$, $W_k$ and $W_q$ are their 
			corresponding effective energies. $\mathbf{P}$ is the center of mass (c.m.) momentum. 
			The Pauli operator     
			\begin{equation}\label{PauliBlockOperator}
			Q_{\tau_1\tau_2}(\mathbf{k},\mathbf{P}) = \left\{
			\begin{array}{cl}
			1  &   \quad  (|\mathbf{P}+\mathbf{k}|>k_F^{\tau_1}~
			\text{and}~|\mathbf{P}-\mathbf{k}|>k_F^{\tau_2}), \\
			0  &   \quad  (\text{otherwise}),
			\end{array}\right.
			\end{equation} 
			can prevent nucleon above the Fermi surface scattering into the Fermi sea in nuclear medium according to the Pauli exclusion principle, where $k_F^\tau$ denotes the Fermi momentum of nucleon $\tau$.
			
			From the Dirac equation in nuclear medium Eq.~\eqref{ImDiracEq}, the expectation value of single-particle potential for nucleon $\tau$ with momentum $\mb{p}$ can be expressed by         
			\begin{equation}\label{singpot-p}
			U_\tau(p)=\frac{M_\tau^*}{E_\tau^*} \langle \mb{p},s|\beta U_\tau|\mb{p},s 
			\rangle=\frac{M_\tau^*}{E_\tau^*}U_{\tau,\mathrm{s}}+U_{\tau,\mathrm{v}}.
			\end{equation} 		
			On the other hand, this single-nucleon potential can be evaluated through the effective $NN$ potential $G$ matrices in the mean-field approximation
			\begin{equation}
			\label{hf-singpot}
			\begin{aligned}
			U_\tau(p) &= \sum_{\tau'}\sum_{ss'}\int^{p' \leqslant k^\tau_F}
			\frac{d^3p'}{(2\pi)^3}\langle \mb{p}s,\mb{p}'s' |G_{\tau\tau'}| 
			\mb{p}s,\mb{p}'s'-\mb{p}'s',\mb{p}s\rangle.
			\end{aligned}
			\end{equation}
			Hence, the scalar and vector potential $U_{\tau,\mathrm{s}}$ and $U_{\tau,\mathrm{v}}$ can be obtained in 
			a self-consistent way by iterative evaluations of Eqs.~\eqref{BbS-Gmatrix}, \eqref{singpot-p}, and
			\eqref{hf-singpot} until their convergence. 
			
			Then the binding energy per nucleon at baryon
			density $n_b=n_p+n_n$ and the asymmetry factor $\alpha=\frac{n_n-n_p}{n_b}$ is evaluated by
			\begin{equation}\label{epera}
			\begin{aligned}
			\overline{E}(n_b, \alpha) &= \frac{1}{n_b}
			\sum_{\tau,s}\int^{p\leqslant k_F^\tau}\frac{d^3 p}{(2\pi)^3}
			\frac{M^*_\tau}{E^*_\tau}\langle \mb{p},s|\bm{\alpha}\cdot \mb{p} + 
			\beta M_\tau |\mb{p},s\rangle -\frac{1-\alpha}{2}M_p - \frac{1+\alpha}{2}M_n\\
			& + \frac{1}{2 n_b} \sum_{\tau\tau'}\sum_{ss'} \int^{p\leqslant k_F^\tau}
			\frac{d^3p}{(2\pi)^3}\int^{p'\leqslant k_F^{\tau'}} \frac{d^3p'}{(2\pi)^3}
			\langle\mb{p}s,\mb{p}'s'|G_{\tau\tau'}|\mb{p}s,\mb{p}'s'-\mb{p}'s',\mb{p}s
			\rangle.
			\end{aligned}
			\end{equation}
			The Eqs.~\eqref{hf-singpot} and~\eqref{epera} are 
			calculated in the nuclear matter rest frame and the $G$ matrix will be decomposed into 
			$|LSJ\rangle$ representation. The explicit solid angles of Pauli operator $Q$ in these integrals are used instead of their averaged values~\citep{alonso2003}, while the angular integrals about the center-of-mass momentum $\mb{P}$ are exactly worked out~\citep{tong2018}.   
		
		\subsection{The neutron star matter}
			In this work, we concentrate on the discussion of the core matter of neutron star, which is regarded as the beta equilibrium nuclear matter with electron ($e$) and muon ($\mu$). The beta equilibrium matter is established on the chemical equilibrium conditions between nucleons and leptons, 
			\begin{equation}\label{chemequilb}
			\mu_l + \mu_p = \mu_n.
			\end{equation} 
			Furthermore, the whole system should be of charge neutrality 
			\begin{equation}\label{chargneut}
			n_p = n_e+n_\mu,
			\end{equation}
			with $\mu_l$, $\mu_p$ and $\mu_n$ being the chemical potentials of 
			leptons, proton, and neutron.			
		
		\subsubsection{The symmetry energy approximation}
			In principle, the chemical potential of particle is obtained by taking the derivative of total energy respect to its number density. However, in the conventional investigations of neutron star with RBHF model, the EOSs of neutron star matter were generally obtained based on the symmetry energy approximation due to the complicated and time-consuming calculation for the neutron-rich matter~\citep{alonso2003,krastev2006,tong2020}. The binding energy per nucleon of asymmetric nuclear matter can be approximately expressed by the binding energy per nucleon of symmetric nuclear matter and symmetry energy as,
			\begin{equation}\label{parab-approx}
			\overline{E}(n_b,\alpha) = \overline{E}_0(n_b)+E_\mathrm{sym}(n_b)\alpha^2,
			\end{equation}
			where, the symmetry energy in RBHF model can be extracted from the energy differences between the symmetric nuclear matter and pure neutron matter,
			\begin{equation}\label{symene}
			E_\mathrm{sym}(n_b) = \overline{E}(n_b,1)-\overline{E}(n_b,0).
			\end{equation}
			Therefore, the total energy per nucleon in neutron star matter with nucleons and leptons can be written as,
			\begin{equation}\label{se-ene}
			\overline{E}_\mathrm{tot}=\overline{E}_0+E_\mathrm{sym}(Y_n-Y_p)^2
			+Y_pM_p+Y_nM_n+\frac{\varepsilon_e}{n_b}+\frac{\varepsilon_\mu}{n_b},
			\end{equation}
			where $Y_i$ is the particle fraction
			\begin{equation}
			Y_i = \frac{n_i}{n_b}\quad (i=n,~p,~l).
			\end{equation}
			According to the thermodynamic definition, the chemical potential for each particle $i$ is  
			\begin{equation}\label{se-chempot}
			\mu_i = \frac{\partial \overline{E}_\mathrm{tot}}{\partial Y_i}.
			\end{equation}
			Through the beta equilibrium condition~\eqref{chemequilb}, the chemical potential of lepton can be related to the symmetry energy at a fixed baryon density.
			\begin{equation}\label{se-chemequilb}
			\mu_l  = M_n-M_p +4(1-2Y_p)E_\mathrm{sym}(n_b).
			\end{equation} 
			Furthermore, the charge neutrality condition~\eqref{chargneut} implies 
			\begin{equation}\label{se-chargneut}
			Y_p=Y_e + Y_\mu.
			\end{equation}		
			After solving Eqs.~\eqref{se-chemequilb} and \eqref{se-chargneut} simultaneously, 
			the proper particle fractions $Y_i$ for nucleons and leptons at a given baryon density $n_b$ can be obtained in neutron star matter. Then the corresponding total energy density $\varepsilon_\text{tot}$ and the pressure $P$ 
			for the beta equilibrium matter can be easily obtained from Eqs.~\eqref{se-ene} and  
			\begin{equation}\label{nm-pre}
			P=-\frac{\partial \overline{E}_\mathrm{tot}}
			{\partial (1/n_b)}=-\frac{\partial (\varepsilon_\text{tot}/n_b)}
			{\partial (1/n_b)}=n_b\frac{\partial \varepsilon_\text{tot}}
			{\partial n_b}-\varepsilon_\text{tot}.
			\end{equation}
			
			\subsubsection{The self-consistent method}
			In this work, the total energy of asymmetric nuclear matter will be directly obtained by solving Eqs.~(\ref{chemequilb}) and (\ref{chargneut}) regularly. The chemical potential of nucleon in mean-field approximation is given by
			\begin{equation}\label{nucl-chmpot}
			\mu_\tau =E_F^\tau =\sqrt{k_F^{\tau 2} + M_\tau^{*2}}+U_{\tau,\mathrm{v}}.
			\end{equation}
			where $U_{\tau,\mathrm{v}}$ is the vector potential, and the effective nucleon mass $M^*_\tau$ is related to the scalar potential. Both of them are dependent on the asymmetry factor $\alpha$. 
			
			The leptons are treated as the non-interacting Fermi gas, whose chemical potential at zero temperature corresponds to its Fermi energy
			\begin{equation}\label{lep-chem}
			\mu_l =E_F^l =\sqrt{k_F^{l2} + m_l^2 }\quad (l=e,~\mu).
			\end{equation} 
			Its energy density is given by
			\begin{equation}\label{lepton-ene}
			\begin{aligned}
			\varepsilon_l=& \frac{1}{\pi^2}\int_0^{k_F^l}dp \cdot p^{2}
			\sqrt{p^2+m_l^2} \\
			=&\frac{k_F^l E_F^{l3}}{4\pi^2}-\frac{k_F^l E_F^l m_l^2}{8\pi^2}
			-\frac{m_l^4}{8\pi^2}\ln \left(\frac{E_F^l+k_F^l}{m_l}\right).  
			\end{aligned}
			\end{equation}	
			
			Finally, once the beta equilibrium and charge neutrality conditions are solved, the energy density of the beta equilibrium matter can be exactly expressed by proton fraction, $Y_p$, 
			\begin{equation}\label{nm-ene}
			\varepsilon_\textrm{tot} = n_b[\overline{E}(n_b,1-2Y_p) + Y_p M_p + (1-Y_p)M_n]
			+\varepsilon_e+\varepsilon_\mu. 
			\end{equation}
			The total pressure can be derived from Eq.~(\ref{nm-pre}) with numerically differential method.
			
		\subsection{The neutron star properties}
			The mass and radius of a cold, spherical, static, and relativistic star, should be described by Tolman-Oppenheimer-Volkov (TOV) equation \citep{tolman1939,oppenheimer1939},
			\begin{equation}\label{ns-toveq}
			\begin{gathered}
			\frac{dP(r)}{dr}=-\frac{[P(r)+\varepsilon(r)][M(r)+4\pi r^3P(r)]}{r[r-2M(r)]}, \\
			\frac{dM(r)}{dr} = 4\pi r^2\varepsilon(r).
			\end{gathered}
			\end{equation}   
		    These differential equations can be solved numerically with a given central pressure, $P_c$ and $M(0)=0$. The $R$ for $P(R)=0$ denotes the radius of neutron star and $M(R)$ its the mass. Recently, with the development of the astronomical observations, another property of neutron star, the tidal deformability, can be extracted from the gravitational wave detectors in the binary neutron star merger~\citep{abbott2018}, which is defined as 
			\begin{equation}\label{dtd}
			\Lambda = \frac{2}{3}k_2 C^{-5}.
			\end{equation}
			Actually, it represents the quadrupole deformation of a compact star in the external gravitational field generated by another compact star. Here $C=M/R$ is the compactness parameter and the second Love number $k_2$~\citep{hinderer2008,hinderer2010} is defined as
			\begin{equation}\label{ns-luv}
			\begin{aligned}
			k_2=&\frac{8C^5}{5}(1-2C)^2[2-y_R+2C(y_R-1)]\times\{6C[2-y_R+C(5y_R-8)] \\
			&+4C^3[13-11y_R+C(3y_R-2) +2C^2(1+y_R)] \\
			&+3(1-2C)^2[2-y_R+2C(y_R-1)]\ln(1-2C)\}^{-1},
			\end{aligned}
			\end{equation}
			where $y_R=y(R)$ is a solution of the following differential equation,   
			\begin{equation}
			r\frac{d y(r)}{dr} + y^2(r)+y(r)F(r) + r^2Q(r)=0.
			\end{equation}
			$F(r)$ and $Q(r)$ are the functions of mass, radius, energy density, and pressure,
			\begin{align}
			F(r) & = \left[1-\frac{2M(r)}{r}\right]^{-1} 
			\left\{1-4\pi r^2[\varepsilon(r)-P(r)]\right\} ,\\ 
			\nonumber 
		    Q(r) & =  \left\{4\pi  \left[5\varepsilon(r)+9P(r)+\frac{\varepsilon(r)
				+P(r)}{\frac{\partial P}{\partial \varepsilon}(r)}\right]-\frac{6}{r^2}\right\}\times 
			\left[1-\frac{2M(r)}{r}\right]^{-1}  \\
			&~~-\left[\frac{2M(r)}{r^2} +2\times4\pi r P(r) \right]^2 \times 
			\left[1-\frac{2M(r)}{r}\right]^{-2} .
			\end{align}
			This differential equation for second Love number $k_2$ can be solved together with TOV equation and the initial condition~$y(0)=2$.

	\section{The results and discussions}\label{results}
 		\subsection{Comparisons between the symmetry energy approximation and self-consistent method}
	 		
\begin{figure}[thb]
	 			\centering
	 			\includegraphics[scale=0.5]{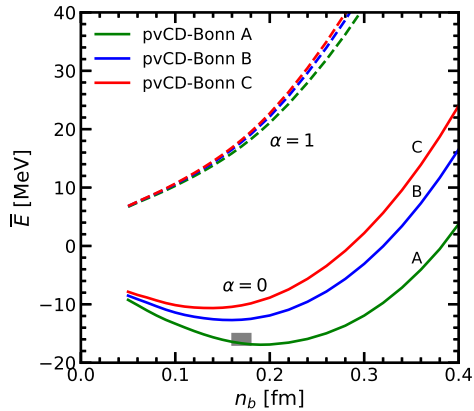}
	 			\caption{The binding energies per nucleon of symmetric nuclear matter ($\alpha=0$) and pure neutron matter ($\alpha=1$) from pvCD-Bonn potentials. The empirical saturation properties of symmetric nuclear matter is shown by the gray block.}
	 			\label{fig-eos}
	 		\end{figure}

	 		\added{First, the binding energies per nucleon of symmetric nuclear matter and pure neutron matter as functions of density from high-precision  pvCD-Bonn A, B, C potentials ~\citep{wang19} are calculated within the RBHF model, which are shown in Fig.~\ref{fig-eos}. The saturation property from pvCD-Bonn A potential mostly approaches the empirical data shown as gray block, which generates the smallest  $D$-state probability and corresponds the weakest tensor component. For pvCD-Bonn A, B, C potentials, the saturation densities and binding energies are $0.192,~0.158, ~0.139$ fm$^{-3}$ and $-16.82,~-12.91,~-10.72$ MeV, respectively, which can form a "Coester band" such as those from Bonn potentials. The equations of state of  pure neutron matter from three pvCD-Bonn potentials are almost identical, since the tensor force plays a negligible role for $T=1$ case.  The symmetry energy of nuclear matter can be approximately expressed  by the differences of the binding energies between the pure neutron matter and symmetric nuclear matter.}
	 			
\begin{figure}[thb]
	 		\centering
	 		\includegraphics[scale=0.5]{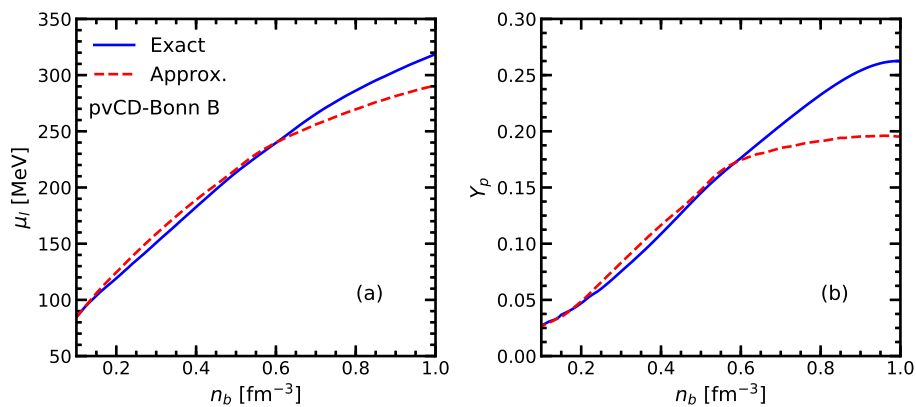}
	 		\caption{The chemical potentials of electron and proton fractions of neutron star matter as a function of baryon density from self-consistent method and symmetry energy approximation. 'Exact' and 'Approx' represent the self-consistent method and symmetry
	 			energy approximation, respectively.
	 			Panel (a) shows results of lepton chemical potentials, while panel (b) shows the corresponding proton fractions.}
	 		\label{fig-chempot}
 		\end{figure}

	 		Then, the results from symmetry energy approximation and self-consistent method in neutron star matter are compared with the same $NN$ potential, pvCD-Bonn B. The BbS equation is solved in the $|LSJ\rangle$ partial wave representation. The total angular momentum $J$ is summed up to $J=8$. In panel~(a) of Fig.~\ref{fig-chempot}, the chemical potentials of electron obtained from these two schemes are shown as functions of baryon density. They are almost identical below $n_b=0.55$ fm$^{-3}$. With the density further increasing, the differences between the two schemes is more obvious. At $n_b=1.0$ fm$^{-3}$, the chemical potential of electron from self-consistent method
	 		is about $320$ MeV, while that from the symmetry energy approximation is $290$~MeV. There is about $10\%$ difference between them.
	 		
	 		In panel~(b) of Fig.~\ref{fig-chempot}, the proton fractions from two schemes are exhibited. 
	 		They show the analogous behaviors to the chemical potentials in panel~(a),
	 		since the chemical potential of lepton is determined by its Fermi momentum, which is related to the proton fraction due to 
	 	    the charge neutrality condition shown in Eq.~\eqref{chargneut}. 
	 		Similarly, above the baryon density $0.55$ fm$^{-3}$, the proton fraction of self-consistent 
	 		method becomes obviously larger than that of symmetry energy approximation. At $n_b=1.00$ fm$^{-3}$, self-consistent method predicts proton fraction as $Y_p=0.26$ while it is about $Y_p=0.20$ in the symmetry energy approximation. There is about $20\%$ difference between two schemes. In addition, the proton fraction from the symmetry energy approximation is approaching a saturation value slowly, which is similar with the results from Krastev and Summarruca~\citep{krastev2006}.  
	 		
	 		To solve the TOV equation, the EOS must cover full regions of neutron star from outer crust to the core. In present work, we just concentrate on the discussion of the core region in neutron star EOS within RBHF model. Therefore, the outer crust applies Baym-Pethick-Sutherland (BPS) EOS~\citep{baym1971} , while the inner 
	 		crust part use the EOS with RMF interaction and self-consistent Thomas-Fermi method
	 		for pasta phase. The RMF interaction is adopted as the TM1 parameterization with symmetry energy slope $L=60$
	 		MeV~\citep{bao2014a,bao2014b}, which is close to those from pvCD-Bonn potentials A, B, C ($L=80,~57,~45$ MeV,
	 		respectively).
	 		
	 		\begin{figure}[thb]
	 			\centering
	 			\includegraphics[scale=0.5]{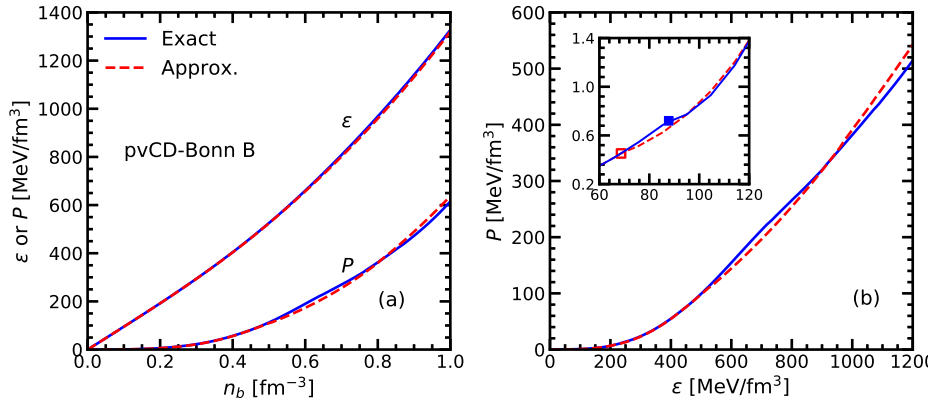}
	 			\caption{Panel~(a): the energy density and pressure as functions of baryon density obtained by 
	 				the self-consistent method and symmetry energy approximation.
	 				Panel~(b): the corresponding EOSs obtained by two schemes. In the insert of panel (b),  
	 				dots indicate the crust-core transition points.}
	 			\label{fig-scse-eos}
	 		\end{figure}
	 		
	 		In panel~(a) of Fig.~\ref{fig-scse-eos}, the energy densities and pressures  as functions of 
	 		baryon density derived from the self-consistent method and symmetry energy approximation are showed, respectively. 
	 		The energy densities given by the two schemes are almost  the same, while the corresponding pressures reveal differences
	 		above the density $0.55$ fm$^{-3}$. It can be easily understood from the thermodynamic self-consistency condition~$\sum_in_i\mu_i=\varepsilon+P$.  In the panel~(b), the pressure as a function of energy density is given. There, the distinction between two methods is more obvious. Furthermore, the crust-core transition density from symmetry energy approximation is smaller than that from the self-consistent method.
	 		
\begin{figure}[thp]
				\centering 
				\includegraphics[scale=0.5]{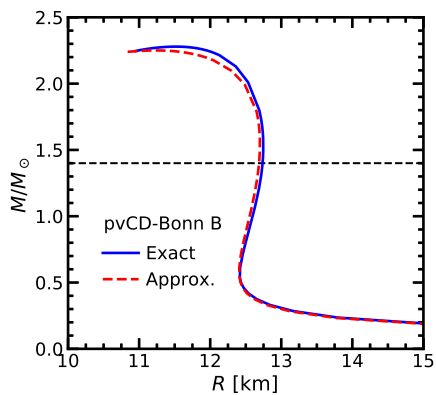}
				\caption{The neutron star mass-radius relations obtained from self-consistent method and symmetry energy approximation.}
				\label{fig-scse-rm}
			\end{figure}

	 		 When these complete EOSs in panel (b) of Fig.~\ref{fig-scse-eos} are applied in the 
	 		 TOV equation,  the mass-radius relation of neutron star can be obtained and plotted in
	 		 Fig.~\ref{fig-scse-rm}. The maximum neutron star mass and the corresponding
	 		 radius are $2.28M_\odot$ and $11.52$ km from self-consistent method, while they are $2.25M_\odot$ and $11.27$ km from symmetry energy approximation. Their mass-radius relations have a few differences above $1.0 M_\odot$. At a fixed radius, the mass of neutron star from the self-consistent method is larger than that from the symmetry energy approximation, since the EOS of former is a bit stiffer in the energy density region $\varepsilon=500\sim950$ MeV/fm$^3$ as shown in panel (b) of Fig.~\ref{fig-scse-eos}.   
	 		 
			In the neutron star cooling process, the direct URCA (DURCA)  reaction 
			\begin{equation}
			n\rightarrow p +e + \bar{\nu}_e, \quad p+e \rightarrow n+\nu_e.
			\end{equation}
			plays a very important role, which requires the conservation of momenta for nucleons and leptons
			$k_F^p + k_F^l =k_F^n$.
			With this equation, the threshold density of DURCA process can be predicted. If muon does not appear, this threshold density corresponds to $Y_p=1/9$~\citep{lattimer1991}. In the calculations of RBHF model, the DURCA process happens after muon's appearance. The threshold densities are predicted as $n_b=0.477$ fm$^{-3}$ and $Y_p=0.137$ from self-consistent method, while the predictions from symmetry approximation are $n_b=0.464$ fm$^{-3}$ and $Y_p=0.136$. It is consistent with the astronomical observations, which does not allow the DURCA process too early.
			
			Finally, the properties of neutron star from two schemes with pvCD-Bonn B potential are listed in Table~\ref{tab-scse}, such as the threshold density of DURCA process, the maximum mass, the central density, the radius, density and tidal deformability at  $1.4M_\odot$. It can be found that the results from symmetry energy approximation are consistent with those from self-consistent method. Their differences are smaller than $3\%$ for these global properties of neutron star. Therefore, the symmetry energy approximation is a very good approach to treat the time-consuming calculations in neutron star. It is worth noting that when we treat the physical processes in the central region of neutron star, especially related to particle fractions, the self-consistent method should be more reliable.  
			
			\begin{table}[h]
				\caption{The neutron star properties obtained form the self-consistent method (Exact) and the symmetry energy approximation (Approx.). 
					$n_{b,\textrm{URCA}}$, $Y_{p,\textrm{URCA}}$ are baryon density and proton 
					fraction respectively for the DURCA process thresholds.
					$R_\textrm{max}$, $n_{b,\textrm{max}}$ are the radius and central baryon 
					density for the neutron star with maximum mass $M_\textrm{max}$.  
					$R_{1.4}$, $n_{b,1.4}$, $k_{2,1.4}$, and $\Lambda_{1.4}$ represent the radius, 
					central baryon density, second Love number, and dimensionless tidal 
					deformability of neutron star with mass 1.4$M_\odot$, respectively. 
					The units of mass, radius and central density are solar mass, km, and 
					fm$^{-3}$.}
				\centering
				\begin{tabular}{r|cc|ccc|cccc}
					\hline 
					\hline  
					&~$n_{b,\textrm{DURCA}}$~& ~$Y_{p,\textrm{DURCA}}$ ~
					& ~$M_\textrm{max}$ ~   & ~$R_\textrm{max}$~ & ~$n_{b,\textrm{max}}$ ~
					&~ $R_{1.4}$ ~& ~$n_{b,1.4}$  ~& ~$k_{2,1.4}$ ~& ~$\Lambda_{1.4}$  ~  \\
					\hline     
					Exact& 0.477 & 0.137 & 2.28 & 11.52 & 0.914 & 12.74 & 0.391 & 0.098 & 580\\ 
					Approx.&0.464 & 0.136 & 2.25 & 11.27 & 0.949 & 12.69 & 0.392 & 0.098 & 578\\
					\hline 
					\hline 
				\end{tabular}
				\label{tab-scse}
			\end{table} 

		\subsection{Neutron star properties from self-consistent calculations with pvCD-Bonn A, B, C potentials}
			In this subsection, the properties of neutron star will be investigated and be compared in the framework of self-consistent method with pvCD-Bonn A, B, C potentials. The most significant difference among these three potentials is their
			strengths of tensor force, which will play an important role in the symmetry energy and influence the proton fractions in neutron star matter. The strengths of tensor force in pvCD-Bonn A, B, C potentials gradually increase with different coupling constants and cutoffs in the one-pion exchange component, similar with the Bonn potentials~\citep{brockmann1990}. The strengths of tensor force can be represented by the $D$-state probability of deuteron, $P_D$, which is strong correlated with the saturation properties of nuclear matter, i.e. the Coester band. In a word, larger $P_D$ indicates stronger tensor force which generates smaller symmetry energy $E_\textrm{sym}$ at given baryon density.
			The $D$-state probabilities, $P_D$ of pvCD-Bonn A, B, C potentials are $4.2\%$, $5.5\%$, and $6.1\%$ respectively.

			\begin{figure}[thb]
				\centering 
				\includegraphics[scale=0.5]{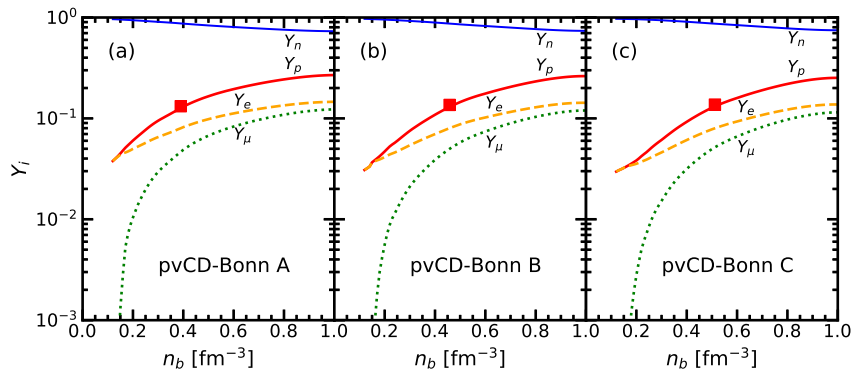}
				\caption{The particle fractions of neutron star matter calculated by self-consistent 
					RBHF model with pvCD-Bonn potentials. 
					The squares on the proton fractions stand for the thresholds of DURCA process.}
				\label{fig-scyp}
			\end{figure}
		
		Based on the Eqs.~\eqref{chemequilb}, \eqref{chargneut} and \eqref{nucl-chmpot},
		the particle fractions, $Y_i$ for beta equilibrium matter can be calculated in RBHF model self-consistently with these
		three pvCD-Bonn potentials, which are plotted as a function of baryon density in Fig.~\ref{fig-scyp}.   
		The onset densities for muon from pvCD-Bonn A, B, C potentials are $0.15$ fm$^{-3}$, $0.17$ fm$^{-3}$
		and $0.18$ fm$^{-3}$ respectively. The muon appears earliest in pvCD-Bonn A potential, because it has the strongest symmetry energy and generates the largest proton fraction.  The threshold densities of DURCA process in neutron star matter, obtained
		from pvCD-Bonn A, B, C potentials are $0.41$ fm$^{-3}$, $0.48$ fm$^{-3}$, and $0.53$ fm$^{-3}$ respectively with the same DURCA proton fractions $Y_p=0.14$, after considering the Fermi momenta conservation of nucleons and leptons.
		Actually, the DURCA threshold density can be simply estimated as be inversely proportional to the symmetry energy
		at given density.

\begin{figure}[h]
				\centering
				\includegraphics[scale=0.5]{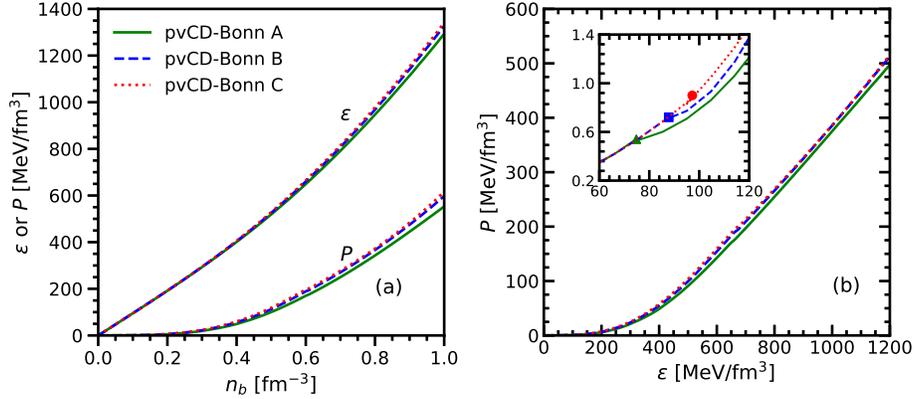}
				\caption{The energy density, pressure as functions of baryon density and the EOS of 
					neutron star  matter plotted in  panels (a) and (b) respectively. 
					Dots in the insert of panel~(b) denote the corresponding crust-core transition points for three pvCD-Bonn potentials.}
				\label{fig-sc-eos}
			\end{figure}

			When the particle fractions are fixed, the pressure and  energy density of neutron star matter can be obtained from 
			Eqs.~\eqref{nm-pre} and \eqref{nm-ene}. They are shown in panel~(a) of Fig.~\ref{fig-sc-eos} as functions of
			baryon density.  There are few differences among them generated from pvCD-Bonn A, B, C potentials since neutron dominates the properties of neutron star matter, while the contribution of tensor force becomes weaker with the isospin asymmetry increasing. Comparatively speaking, their differences in pressure are more obvious at high density. The EOS from pvCD-Bonn A potential is softest with the smallest component of tensor force. In panel~(b), the energy densities as functions of pressure are shown. As mentioned before, in present work, the RBHF model is  only applied to calculate the uniform matter in the core part. In low density region, we adopt the BPS EOS and that from self-consistent Thomas-Fermi method. The matching points of the crust and core EOSs are chosen where both the energy densities and pressures are same. The corresponding crust-core transition densities  are $0.077$ fm$^{-3}$, $0.092$ fm$^{-3}$, and $0.100$ fm$^{-3}$ respectively for pvCD-Bonn A, B, C potentials.
		
\begin{figure}[thb]
				\centering 
				\includegraphics[scale=0.5]{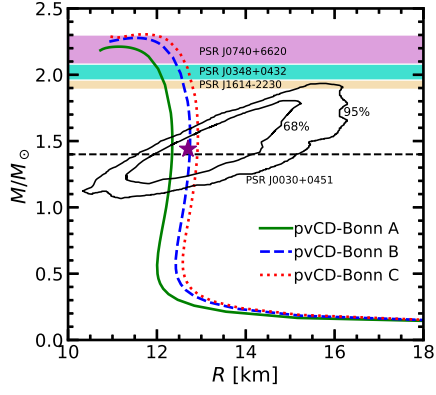}
				\caption{The mass-radius relations from pvCD-Bonn A, B, C potentials.  
					The shaded areas are the constraints from the massive neutron star observations, taken from Refs.
					\citep{cromartie2020, antoniadis2013,demorest2010, fonseca2016}.
					The inner and outer contours indicate the $68\%$ and $95\%$ confidence intervals constrained by NICER's analysis of PSR J0030+451 \citep{miller2019}.}
				\label{fig-sc-rm}
			\end{figure}

			Within these EOSs, the mass-radius relations of neutron star can be generated from the TOV equation in Eq.~\eqref{ns-toveq}. In Fig.~\ref{fig-sc-rm}, the maximum masses of neutron star are predicted as $2.21M_\odot$, $2.28M_\odot$,
			$2.30M_\odot$ from the pvCD-Bonn A, B, C potentials respectively, which are consistent with the available observations about  massive neutron stars, such as PSR J1614-2230, PSR J0348+0432, and PSR J0740+6620. The stiffer EOS leads to a larger maximum neutron star mass, therefore pvCD-Bonn C potential produces the heaviest neutron star. The corresponding radius are $11.18$, $11.54$, and $11.72$ km. Due to the different strengths of tensor force in pvCD-Bonn potentials, the core densities of neutron star at maximum mass are quite distinguished.
			It is $0.97$ fm$^{-3}$ for pvCD-Bonn A potential and is about $10\%$ larger comparing to the one from pvCD-Bonn C potential. \added{Therefore, these three mass-radius curves from pvCD-Bonn potentials at large mass region have obvious differences due to their tensor components. In the lower mass region, these relations are quite different, where the proton fractions are rather small. This is because the crust EOSs from Thomas-Fermi method are adopted in the low density region, and the crust-core transition densities have some differences as shown in Fig.~\ref{fig-sc-eos}. }
			
			Recently, the mass and radius of PSR J0030+451 were observed simultaneously by NICER. Miller et al. estimated that its mass is $1.44^{+0.15}_{-0.14}M_\odot$ with radius $13.02^{+1.24}_{-1.06}$ km~\citep{miller2019}. In present calculations from RBHF model, the radii at $1.4M_\odot$ neutron star are $12.34$ km (pvCD-Bonn A), $12.77$ km (pvCD-Bonn B), $12.91$ km (pvCD-Bonn C), which are completely consistent with the constraint from NICER. The confidence intervals for $68\%$ and $95\%$ about the relations between mass and radius from the NICER analysis are also shown. It can be found that the results from three pvCD-Bonn potentials are properly located in these confidence intervals, especially those from pvCD-Bonn B potential, which completely coincide with the central values of the NICER analysis data shown as the star symbol in Fig~\ref{fig-sc-rm}. 

\begin{figure}[thb]
				\centering 
				\includegraphics[scale=0.5]{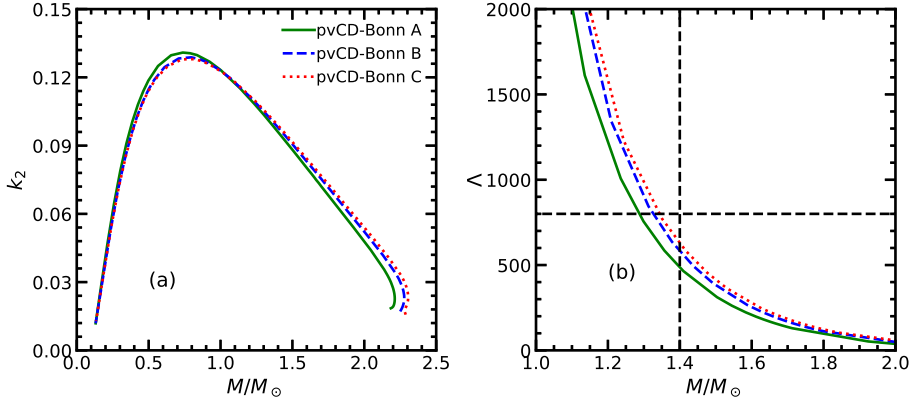}
				\caption{The second Love number and dimensionless deformabilities obtained by RBHF model within
						 pvCD-Bonn A, B, C potentials.}
				\label{fig-SC-tidal}
			\end{figure}

			In 2017, the gravitation wave from the binary neutron star merger was detected by advanced LIOG and Virgo collaborations as GW 170817 event, which provided new constraints for the tidal deformabilities of neutron star at intermediate neutron star mass region. The tidal deformability represents the quadrupole deformation of a compact star in an external gravitational field from another star within a binary star system, which is related to the second Love number $k_2$. This $k_2$ and dimensionless
			tidal deformability $\Lambda$ from pvCD-Bonn potentials are shown in Fig.~\ref{fig-SC-tidal}. 
			Panel (a) presents the second Love number $k_2$ as a function of the neutron star mass. It firstly increases with neutron star mass, and reaches its maximum value around $0.13$ at $M=0.8M_\odot$, the rapidly reduces at large mass region. The second Love number $k_2$ at $1.4M_\odot$ from pvCD-Bonn A, B, C potentials are $0.096$, $0.098$, and $0.099$, respectively. 			
			 
			The dimensionless tidal deformabilities $\Lambda$ as functions of neutron star mass are plotted in the panel (b) of Fig.~\ref{fig-SC-tidal}. It decreases with neutron star mass dramatically, since it is strongly dependent on the compactness parameter $C=M/R$ as shown in Eq. (\ref{dtd}). The initial estimation for $\Lambda$ at $1.4 M_\odot$ was less than $800$ from GW 170817. The revised analysis from LIGO and Virgo collaborations showed $\Lambda_{1.4}=190^{+390}_{-120}$~\citep{abbott2018}. Furthermore, there are also many works to constrain the $\Lambda_{1.4}$ with the observational data from the gravitational wave detector. The values of $\Lambda_{1.4}$ are $485, 580$, and $626$ from pvCD-Bonn A, B, C potentials, respectively, which are very similar with results from Bonn A, B, C potentials by Tong et al.~\citep{tong2020} and are consistent with the constraint of gravitational wave. 
			
\begin{figure}[h]
				\centering 
				\includegraphics[scale=0.5]{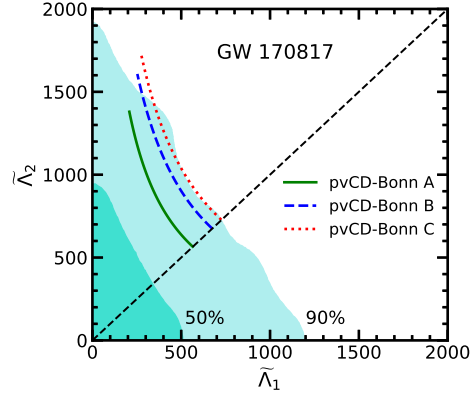}
				\caption{The tidal deformabilities of the binary components in GW 170817 with the constraints from the analysis of GW 170817 \citep{abbott2018}. $\Lambda_1$ corresponds to the larger mass component in binary system.}
				\label{fig-td170817}
			\end{figure}

			In present gravitational wave detection, it is very difficult to distinguish the corresponding masses of two neutron stars in merger process, while their chirp mass, defined as $\mathcal{M}=(M_1M_2)^\frac{3}{5}(M_1+M_2)^{-\frac{1}{5}}$, 
			can be exactly extracted from the gravitational wave signal. The chirp mass in GW 170817 was measured as $\mathcal{M}=1.188^{+0.004}_{-0.002}M_\odot$~\citep{abbott2017a}. Therefore, if we assume that the mass of one neutron star is in the range from $1.170M_\odot$ to $1.365M_\odot$, while the other star has $1.365M_\odot$ to $1.600M_\odot$ with the constraint of chirp mass. The corresponding dimensionless tidal deformabilities are named as $\Lambda_2$ and $\Lambda_1$.  Their confidence intervals for $50\%$ and $90\%$ from GW 170817 observations are plotted as the shadow areas in Fig.~\ref{fig-td170817}. The correlation between $\Lambda_1$ and $\Lambda_2$ calculated by the RBHF model with different pvCD-Bonn potentials are completely located within these constraints. 

\begin{figure}[h]
				\centering
				\includegraphics[scale=0.5]{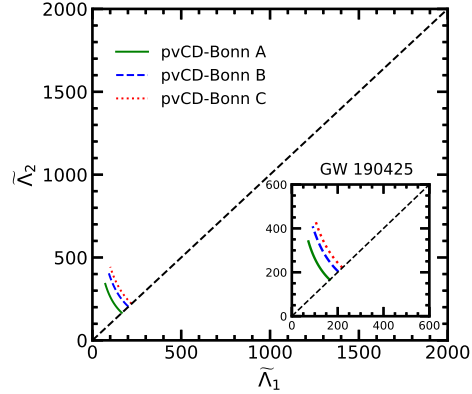}
				\caption{The tidal deformabilities predictions for two neutron star components in GW 190425 event~\citep{abbott2020} .}
				\label{fig-td190425}
			\end{figure}

			In addition to the event GW 170817, another gravitational wave event, GW 190425, was considered as one possible binary neutron star merger \citep{pozanenko2019,abbott2020}, while it is also possible a neutron-black hole merger. The total mass of the binary system in GW 190425 is around $3.4M_\odot$, with the chirp mass $\mathcal{M}=1.44_{-0.02}^{+0.02}~M_\odot$. If both components in this binary system are regarded as neutron stars, according to results of $M$--$\Lambda$
			relation in present framework, the joint tidal deformabilities for each components can be predicted in Fig.~\ref{fig-td190425}. It can be found that the tidal deformabilities in GW 190425 are much smaller than those in GW 170817 since the neutron star masses of the former are larger.

				Finally, the numerical details of neutron star properties, such as threshold of DURCA process, maximum mass, central density, and the radii and tidal deformability at $1.4M_\odot$,  obtained by pvCD-Bonn A, B, C potentials together with Bonn A, B, C potentials in the framework of self-consistent method are collected in Table ~\ref{tab-sc}.  In DURCA process, it can be found that the threshold densities are around $0.414-0.530$ fm$^{-3}$. The maximum neutron star masses are in the region $2.21-2.30M_\odot$ and corresponding radii $11.18-11.72$ km. The radii of $1.4M_\odot$ locate in $12.34-12.91$ km, which is consistent with the recent constraints from various observations. The corresponding tidal deformabilities are $485-626$. Actually, it is easily found that these properties of neutron star are strongly correlated to the tensor components of $NN$ potentials with the systematical calculations.
			
\begin{figure}[h]
					\centering 
					\includegraphics[scale=0.5]{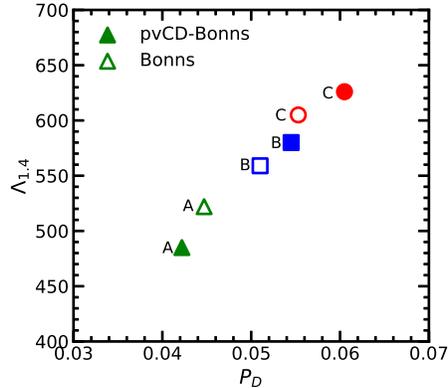}
					\caption{Dimensionless tidal deformabilities of $1.4M_\odot$ neutron star, $\Lambda_{1.4}$ predicted by different $NN$ potentials with their deuteron $D-$state probabilities, $P_D$.}
					\label{fig-sc-pd}
				\end{figure}

			    \replaced{For example, in Fig.~\ref{fig-sc-pd}, the tidal deformabilities of $1.4M_\odot$ from the pvCD-Bonn and Bonn potentials almost have the linear correlation with the $D-$state probabilities in deuteron, $P_D$, which is very analogous to the "Coester band" for the saturation density in symmetric nuclear matter.}{The tensor component of $NN$ potential usually can be denoted by the $D$-state probability of deuteron, $P_D$. The larger value of $P_D$ corresponds to the stronger tensor force. In the available {\it ab initio} calculations~\citep{brockmann1990,li06} in nuclear matter, it was found the saturation properties of symmetric nuclear matter from different $NN$ potentials have the linear correlations with their $D-$state probabilities, namely the larger $P_D$ potential generated smaller saturation density and larger saturation binding energy. It is also called as 'Coester band'. In Fig.~\ref{fig-sc-pd}, the  tidal deformabilities of $1.4M_\odot$ from the pvCD-Bonn and Bonn potentials are shown as a function of their $D-$state probabilities. It is clear to see that a potential with larger $P_D$ leads to a larger tidal deformability. Furthermore, there is also a linear correlation between  $D-$state probability of $NN$ potential and the tidal deformability of neutron star, since the tensor force still provides some contributions in neutron star matter, where the proton fraction is about $10\% -20\%$. }
			    
				\begin{table}[h]
					\centering 
					\caption{The neutron star properties from pvCD-Bonn and Bonn potentials.}
					\begin{tabular}{r|cc|ccc|cccc}
						\hline 
						\hline  
						& ~$n_{b,\textrm{DURCA}}$~& ~$Y_{p,\textrm{DURCA}}$ ~
						& ~$M_\textrm{max}$  ~   &~ $R_\textrm{max}$ ~& ~ $n_{b,\textrm{max}}$ ~
						& ~$R_{1.4}$ ~&~  $n_{b,1.4}$ ~ & ~$k_{2,1.4}$~ & ~$\Lambda_{1.4}$   ~  \\
						\hline     
						pvCD-Bonn A&0.414&0.136&2.21 & 11.18  & 0.970 &12.34& 0.425 & 0.096&485\\
						pvCD-Bonn B&0.477&0.137&2.28 & 11.54  & 0.921 &12.77& 0.392 & 0.098&580\\ 
						pvCD-Bonn C&0.530&0.138&2.30 & 11.72  & 0.880 &12.91& 0.376 & 0.099&626\\
						\hline 
						Bonn A&0.416&0.135&2.22 & 11.29  & 0.950 &12.48& 0.412 & 0.092&522\\
						Bonn B&0.463&0.136&2.22 & 11.36  & 0.941 &12.58& 0.403 & 0.100&559\\
						Bonn C&0.514&0.137&2.23 & 11.42  & 0.932 &12.63& 0.397 & 0.110&605\\
						\hline 
						\hline 
					\end{tabular}
					\label{tab-sc}
				\end{table} 

	\section{Summaries and perspectives}\label{summary}
	    A relativistic {\it ab initio} method, i.e., the relativistic Brueckner-Hartree-Fock (RBHF) model, was applied to study the properties of neutron star with the latest relativistic high-precision nucleon-nucleon ($NN$) potentials, pvCD-Bonn A, B, C. These three potentials can completely describe the $NN$ scattering phase shifts and include different components of tensor force. The neutron star matter in present framework was considered as the compositions of protons, neutrons, electrons, and muons with the beta equilibrium and charge neutrality conditions.
	    
	    Due to the complication and time-consuming of RBHF model for the asymmetric nuclear matter, in the past available investigations of neutron star within RBHF model, the total energy of neutron star matter was approximately given by the binding energy of symmetric nuclear matter and symmetry energy, named as symmetry energy approximation. In present framework, the equations of state of neutron star matter were solved under the beta equilibrium and charge neutrality conditions self-consistently. It was found that the global properties of neutron star, such as maximum mass, radius, tidal deformability, from two schemes are almost identical for pvCD-Bonn B potential. Their differences appeared at the high density region above $n_b=0.55$ fm$^{-3}$, especially for particle fractions. There are  $20\%$ differences for proton fraction at $n_b=1.0$ fm$^{-3}$ between symmetry energy approximation and self-consistent method. It demonstrates that the symmetry energy approximation is a very good scheme to describe the global properties of neutron star in RBHF model. When we discuss the processes in the central region of neutron star, particle fractions from the self-consistent method are preferred.

		Finally, the properties of neutron star were calculated within pvCD-Bonn A, B, C potentials and self-consistent method.
		Their different tensor components had a few influences on the equations of state of neutron star matter and properties of neutron star, because there were still about $10\%-20\%$ protons existing in neutron star. The maximum neutron star masses from three pvCD-Bonn potentials are $2.21-2.30M_\odot$ and the corresponding radius are $11.18-11.72$ km. The DURCA threshold densities in neutron star cooling are predicted as $0.414-0.530$ fm$^{-3}$. The radius of $1.4M_\odot$ are around $12.34-12.91$ km, where the tidal deformabilities are $485-626$. All of these results completely agree with recent various observations about neutron star, such as massive neutron star observations, LIGO and Virgo detection for the gravitational wave from binary neutron star merger, and the simultaneous measurement for the mass and radius of neutron star from NICER. Furthermore, it was found that the properties of neutron star have strongly correlated to the tensor components of $NN$ potentials, which can generate the "Coester band" as a function of the $D$-state probability of deuteron, $P_D$.
		
		 It fully demonstrated that the RBHF model with the high-precision $NN$ potentials is a very powerful theoretical framework for the astrophysics, which can exactly describe the global properties of neutron star by using only two-body potentials. We will apply this method to study the crust region of neutron star to realize the theoretical self-consistency for the equation of state of neutron star matter. The neutron superfluidity and cooling process in neutron star will also be investigated in future.   

	\section{Acknowledgments}
This work was supported in part by the National Natural Science Foundation of China (Grants  No. 11775119, No. 11675083, and No. 11405116).  the Natural Science Foundation of Tianjin, and China Scholarship Council (Grant No. 201906205013 and No. 201906255002).


\listofchanges 
\end{document}